\newlength{\extralineskip}
\documentclass[12pt]{article}

\begin{document}
\begin{titlepage}
\begin{flushright}
          \begin{minipage}[t]{12em}
          \large UAB--FT--420\\
                 October 1997
          \end{minipage}
\end{flushright}
\vspace{\fill}

\vspace{\fill}

\begin{center}
\baselineskip=2.5em

{\large \bf CONTACT INTERACTIONS INVOLVING RIGHT-HANDED
NEUTRINOS AND SN1987A}
\end{center}

\vspace{\fill}

\begin{center}
{\bf J.A. Grifols, E. Mass\'o, and R. Toldr\`a}\\
\vspace{0.4cm}
     {\em Grup de F\'\i sica Te\`orica and Institut de F\'\i sica
              d'Altes Energies\\
     Universitat Aut\`onoma de Barcelona\\
     08193 Bellaterra, Barcelona, Spain}
\end{center}
\vspace{\fill}

\begin{center}
\large Abstract
\end{center}
\begin{center}
\begin{minipage}[t]{36em}
We consider lepton-quark contact interactions in models with
right-handed neutrinos, and find that observational data from
SN1987A restricts the scale of such interactions to be 
at least $\Lambda>90$ TeV.
\end{minipage}
\end{center}

\vspace{\fill}

\end{titlepage}

\clearpage

\addtolength{\baselineskip}{\extralineskip}

Nonstandard lepton-quark contact interactions 
may arise at low energies
as a consequence of a common quark and lepton substructure or
of heavy boson exchanges, when integrated out. There has
recently been a renewed interest~\cite{general} in contact
interactions due to the fact that they are one of the potential
explanations of the excess of events reported by 
H1~\cite{h1} and ZEUS~\cite{zeus}, at HERA, when measuring
deep-inelastic $e^+ p$ scattering at high-$Q^2$ and comparing
with Standard Model predictions.

The four-fermion operators that could contribute to the HERA 
excess involve the electron and the $u$ and $d$ quarks. The 
corresponding Lagrangian is written as an effective 
electron-quark interaction 
\begin{equation}\label{eq}
{\cal L}_{eq} \; = \;
\sum_{\scriptstyle I,J=L,R \atop \scriptstyle q=u,d}\;
\widetilde\eta^q_{IJ}\,
\frac{4\pi}{(\widetilde\Lambda^q_{IJ})^2}\,
\bar{e}_I \gamma^\mu e_I\,
\bar{q}_J \gamma_\mu q_J\; .
\end{equation}
The factors $\widetilde\eta$ can be $+1$ or $-1$, allowing for
constructive or destructive interferences. In Eq.~(\ref{eq}) 
the various high-energy scales $\widetilde\Lambda$ could be quite
different, so that one allows for
the possibility that only a subset of terms 
in ${\cal L}_{eq }$ is relevant. It is clear that if one 
of the scales is much lower than
the rest of scales, the corresponding term will be the 
dominant at low energies. Of course, it may very well 
be that the dominant
scales are two, or more, because they are of the same
order of magnitude. Also it could happen that some energy 
scales $\widetilde\Lambda$ are equal but with a combination of 
$\widetilde\eta$ factors such that there are cancellations
among them. Such cancellations, in order to be
natural, should occur because of underlying symmetries.

Restrictions on the terms in Eq.~(\ref{eq}) have been
elaborated in the past~\cite{PDG} and now 
reexamined~\cite{general} at the light of the HERA
results. The constraints come from high-energy
accelerators and from precision experiments. 
If the HERA anomaly is going to persist
and the explanation comes indeed from nonstandard
contact interactions, it follows from those 
studies that some particular combinations
of chiralities in Eq.~(\ref{eq}) are preferred 
over others, and also that some cancellations 
should occur in order not to be in conflict 
with precision experiments as atomic parity violation
observations~\cite{general}. The scale that fits 
the HERA data and is compatible with other   
experimental constraints turns out to be on the
range $\widetilde\Lambda \sim 3 - 4$ TeV.

It is worth to study all implications of
contact interactions. With additional hypotheses, one is
able to find further restrictions. 
For example, if such interactions
come from exchanges of heavy bosons, and there are 
bosons with electric charge, one expects charged contact 
interactions~\cite{cornet}. Another aspect that has been
worked out is the fact that gauge-invariance implies
the presence of other operators with the same 
strengths~\cite{deshpande}. Also, the implications
of universality have been studied~\cite{bw}.   

Although HERA directly probes electron-quark contact interactions,
it is clear that lepton-quark interactions need not be restricted
to charged leptons (e.g. electrons) only. 
In the present letter we will consider lepton-quark contact 
interactions in models where a right-handed
neutrino is present. The right-handed neutrino $\nu_R$ appears
quite generally in any extension of the SM.
It is natural that neutrinos
participate in the contact lepton-quark interactions that
would have then a structure similar to Eq.~(\ref{eq})
\begin{equation}\label{nuq}
{\cal L}_{\nu q} \; = \;
\sum_{\scriptstyle I,J=L,R \atop \scriptstyle q=u,d}\;
{\eta}^q_{IJ}\,
\frac{4\pi}{({\Lambda}^q_{IJ})^2}\,
\bar{\nu}_I \gamma^\mu \nu_I\,
\bar{q}_J \gamma_\mu q_J\; . 
\end{equation}
We will focus our attention on the electronic $\nu$, 
so that both Lagrangians Eqs.~(\ref{eq}) and~(\ref{nuq}) 
refer to the fermions in the first generation. However, for
the ease of notation we will not display the subscript ``$e$''
when writing $\nu$. We assume that $\nu_R$ in Eq.~(\ref{nuq}) is a
singlet of $SU(2)_L$ and, together with its left-handed partner 
constitutes the four-component electron neutrino with (Dirac) mass
$m_{\nu_e} <  10-15$ eV \cite{PDG}.

A priori, one expects that the dominant scale $\Lambda$ in the neutrino
sector is of the same order of magnitude as the dominant scale 
$\widetilde\Lambda$ in the electron sector. Should this be the case, then 
Atomic Parity Violation (APV) experiments impose severe limits on the scale 
$\Lambda$. Indeed, APV in cesium \cite{noecker} requires $\widetilde\Lambda > 
10$ TeV and, hence,
$\Lambda > 10$ TeV for the dominant interaction in Eq.~(\ref{nuq}). To elude
this bound (and, hence, to comply with the HERA requirement $\widetilde\Lambda 
\sim 3-4$ TeV) one demands that the Lagrangian Eq.~(\ref{eq}) is parity 
conserving (i.e. left and right couplings are equal). In case {\bf B} below
we consider a kind of scenario for our effective Lagrangian 
Eq.~(\ref{nuq}) where we assume that the underlying theory is vector like.
However, the requirement $\Lambda \sim \widetilde\Lambda$ can be obviously
relaxed in a general phenomenological analysis, and this we do in case 
{\bf A} below where no extra relationships arising from underlying 
left-right symmetries are being imposed. Note that this procedure 
does not conflict APV results since the interaction in Eq.~(\ref{nuq}) do not 
participate in APV effects. 
 
The Lagrangian (\ref{nuq}) allows for emission of the right-handed 
electron neutrinos (and 
left-handed antineutrinos) by the dense nuclear medium in the core of a   
collapsing star as long as these neutrinos are light, $m_\nu \ll T$, where
$T\simeq 50$ MeV is the core temperature. 
The purpose of the present letter is
to show that contact interactions involving right-handed
neutrinos have potential effects in a supernova and to use the
Supernova 1987A observational data to restrict the such hypothetical 
interactions. It has become a standard
procedure to use observational data to limit exotic effects 
affecting stellar energy losses~\cite{raffelt}.
These contact-interaction mediated processes constitute a new channel 
of energy 
drain in the nascent neutron star and therefore may alter its standard  
evolution. If a large amount of the gravitational energy released in the 
collapse escaped the star as 
a flux of right-handed neutrinos, the duration of the ordinary neutrino burst  
would be significantly affected. The observation of  
a neutrino burst in terrestrial  
underground detectors coming from SN1987A in the Large Magellanic 
Cloud sets constraints on the flux of $\nu_R$ produced in its core,  
and allows us to obtain bounds on the coupling constants in Eq.~(\ref{nuq}). 

The part in Eq.~(\ref{nuq}) involving the right-handed neutrino can 
be written as
\begin{eqnarray} \label{lagr3} 
{\cal L}_R &=& \sum_{q=u,d} 2 \pi \left(  
  \frac{\eta^q_{RL}}{(\Lambda^q_{RL})^2}+ 
  \frac{\eta^q_{RR}}{(\Lambda^q_{RR})^2} \right)  
  \bar{\nu_R} \gamma^\mu \nu_R\, \bar{q} \gamma_\mu q \nonumber \\&+& 
               \sum_{q=u,d} 2 \pi \left( 
  -\frac{\eta^q_{RL}}{(\Lambda^q_{RL})^2}+ 
  \frac{\eta^q_{RR}}{(\Lambda^q_{RR})^2} \right) 
  \bar{\nu_R} \gamma^\mu \nu_R\, \bar{q} \gamma_\mu \gamma_5 q \; .   
\end{eqnarray} 
 
It is natural to suppose that the two scales $\Lambda_{RL}$ and 
$\Lambda_{RR}$ are roughly similar. However, the signs $\eta_{RL}$ 
and $\eta_{RR}$ can introduce  
cancellations arising from underlying symmetries of the theory. 
In order to obtain a bound on the energy scales $\Lambda$ one has to  
distinguish two possible cases. For the first case, 
that we call {\bf A}, we suppose 
there are no cancellations so that the vectorial and the axial coupling 
of quarks to right-handed neutrinos are both present, with roughly the same  
strength. In the second case, that we call {\bf B}, we accept 
the following relationship to be true  
\begin{equation} \label{caseB}
\frac{\eta^q_{RL}}{(\Lambda^q_{RL})^2} = 
\frac{\eta^q_{RR}}{(\Lambda^q_{RR})^2} \; , 
\end{equation} 
hence, only the vectorial coupling to quarks is present. 
Let us examine each case separately. 
 
Case {\bf A}. When the vectorial and axial couplings of quarks to neutrinos 
are both present in the Lagrangian with similar weight, i.e. similar  
coupling constant, the problem is analogous to that of emission of ordinary  
neutrinos, which interact with the nuclear medium by means of the effective 
Fermi Lagrangian. The main production process 
is bremsstrahlung of neutrino pairs by the interacting nucleons of the  
medium. It can be shown that the axial coupling dominates over the  
vectorial one~\cite{friman}. In the axial case the source of neutrinos 
is the time fluctuating nucleon spin density, while in the vectorial 
case the source of neutrinos would be the time fluctuations in the 
nucleon number density, which remains constant to a good 
approximation~\cite{raffelt}. 
The total energy carried off by neutrinos per unit time and volume is  
calculated in~\cite{friman}. These authors describe the nucleon 
interaction using 
the one pion exchange~(OPE) approximation and perform the calculation 
in two different extreme cases for the nuclear medium:  
nondegenerated nucleons~($ND$) and extremely degenerated nucleons~($D$).  
To adapt their results to our case we only have to make the replacement: 
\begin{equation} 
\sqrt{2} C^N_A G_F \rightarrow \frac{4\pi}{\Lambda^2}\; , 
\end{equation} 
with $C^N_A$ the axial coupling between nucleons and ordinary neutrinos 
and $G_F$ the Fermi constant. We have defined
\begin{equation} 
\frac{4\pi}{\Lambda^2} \equiv 2\pi  
                \left| -\frac{\eta^q_{RL}}{(\Lambda^q_{RL})^2}+  
                \frac{\eta^q_{RR}}{(\Lambda^q_{RR})^2}   \right|\; ,
\end{equation}
where we consider the dominant quark contribution and drop the upperindex $q$.
In this fashion we obtain (neglecting the pion mass) 
\begin{eqnarray} 
Q_{\nu \bar{\nu}}^{ND} & = & \frac{16384}{385\pi^{3/2}} \ 
  \frac{\alpha_\pi^2}{m^2_N} \ \frac{1}{\Lambda^4} \ n_n n_p \ 
  \left( \frac{T}{m_N} \right)^{1/2} T^5, \\    
Q_{\nu \bar{\nu}}^{D} & = & \frac{328\pi^3}{4725} \ \alpha_\pi^2 
  \ \frac{1}{\Lambda^4} \ p_F \ T^8,
\end{eqnarray}
being $\alpha_\pi \approx 30$ the strong coupling constant for the 
vertex $np\pi$, $m_N\approx 939$ MeV the nucleon mass, $n_n$ and $n_p$
the neutron and proton number densities, respectively, $T$ the core 
temperature and $p_F$ the Fermi momentum of the nucleons.
The total energy carried off by $\nu_R$ and $\bar{\nu}_L$ per unit time  
is $Q_{\nu \bar{\nu}} V_c$, where $V_c$ is the core volume of the 
collapsing star. It cannot exceed $10^{52}$~erg/s, the gravitational
power released by SN1987A in the form of standard neutrinos. This 
observational constraint renders the following bounds on $\Lambda$: 
\begin{eqnarray} 
\Lambda & > & 170 \; \mbox{TeV}   \; \; \; \; \mbox{ND nucleons}, \\ 
\Lambda & > & 250 \; \mbox{TeV}  \; \; \; \; \mbox{D nucleons}. 
\end{eqnarray} 
We have used the standard supernova parameters $n_n = 7\times 10^{38}$  
cm$^{-3}$, $n_p = 3\times 10^{38}$ cm$^{-3}$, $T = 50$ MeV and $R_c = 10$ km. 
In the real situation the nuclear medium is neither nondegenerate nor  
degenerate but in between. The true bound on $\Lambda$ falls then between 
these two limiting cases\footnote{It is interesting to point out that, 
at least for axion bremsstrahlung, the nondegenerate calculation seems  
to be a better approximation when compared with numerical  
calculations~\cite{brinkmann}.}. 
 
Case {\bf B}. Now the dominant axial coupling to nucleons is absent,
since the relation (\ref{caseB}) holds. 
Only the vectorial coupling is left. As discussed above, one expects that  
neutrino bremsstrahlung by nucleons is small. In the absence of the axial  
coupling another process should be considered: emission of neutrino pairs 
by the virtual pions exchanged by the nucleons. This sort of process has 
never been considered previously in the literature, and although one expects 
an emission rate smaller than that found in case {\bf A}, it is now 
interesting to study its contribution to the energy drain.  
 
The Lagrangian (\ref{lagr3}) induces the following structure
\begin{equation}\label{LB} 
\langle \pi | {\cal L}_R | \pi \rangle \, = \, \sum_{q=u,d}
\frac{4\pi}{\Lambda^2}\, \bar{\nu_R} \gamma^\mu \nu_R \,  
      \langle \pi | \bar{q} \gamma_\mu q |\pi \rangle \; , 
\end{equation}
where now
\begin{equation} 
\frac{4\pi}{\Lambda^2} \equiv 2\pi  
                \left| \frac{\eta^q_{RL}}{(\Lambda^q_{RL})^2}+  
                \frac{\eta^q_{RR}}{(\Lambda^q_{RR})^2}   \right| \; .
\end{equation}
Using Lorentz invariance one can write the matrix element in
(\ref{LB}) as
\begin{equation} 
\langle \pi (p') | \bar{q} \gamma_\mu q |\pi (p) \rangle = 
  A(p,p') \left( p+p' \right)^\mu+ B(p,p') \left( p-p' \right)^\mu. 
\end{equation} 

When contracted with the neutrino current the term proportional to $B$ 
is negligible for $m_\nu \ll T$. The function $A$ is, to lowest order, 
1, -1, 0 (quark $u$) and -1, 1, 0 (quark $d$) for the pions 
$\pi^+$, $\pi^-$, $\pi^0$, respectively. Therefore, the only process that 
has to be considered is 
\begin{equation}
n(p_1)\ p(p_2) \rightarrow n(p_4)\ p(p_3) \ \nu (q) \ \bar{\nu}(q')\; .
\end{equation}  
Using the OPE approximation, the squared amplitude summed over spins can 
be written, in the appropriate 
nonrelativistic limit for the nucleons, as 
\begin{equation} 
\sum_{\scriptstyle spin} |{\cal M}(np\rightarrow np \nu \bar{\nu}) |^2 
  = \frac{1}{\Lambda^4} M^{ij}N^{ij}, 
\end{equation} 
where 
\begin{eqnarray} 
M^{ij} & = & 4 \alpha^2_\pi \frac{p^4}{(p^2+m_\pi^2)^4} p^i p^j , \\ 
N^{ij} & = & 8 \left(  q^i q'^j + q^j q'^i + (qq')\delta^{ij} \right),  
\end{eqnarray}
being $\vec{p}$ the three momentum exchanged by the nucleons and $m_\pi$
the charged pion mass.
This factorization is reminiscent of the factorization in a nuclear 
form factor and an emission term that appears when studying bremsstrahlung 
of neutrinos or axions by nucleons~\cite{raffelt}. However, it is crucial 
to realize that the mentioned nuclear form factor does not coincide with  
the present ``nuclear form factor'' $M^{ij}$, since now we have an additional 
pion propagator stemming from the neutrino emission by the virtual pion. 
In the nonrelativistic limit one can write 
\begin{equation} 
Q_{\nu \bar{\nu}} = \frac{1}{20\pi^4} \frac{1}{\Lambda^4} \int_0^{\infty} 
  d\omega \ \omega^6 q(\omega) ,
\end{equation} 
being 
\begin{eqnarray} \label{phspace}
q(\omega) & \equiv & \frac{4\alpha_\pi^2}{3} (4\pi)^4 \int  \prod_{i=1}^4  
  \frac{d\vec{p}_i}{2m_N (2\pi)^3} \ f_1 f_2 (1-f_3) (1-f_4) \nonumber \\ 
  & &\delta^3(\vec{p_1}+\vec{p_2}-\vec{p_3}-\vec{p_4}) \;
  \delta\left(\frac{p_1^2+p_2^2-p_3^2-p_4^2}{2m_N}-\omega\right) 
  \ \frac{p^6}{(p^2+m_\pi^2)^4} 
\end{eqnarray} 
with $\vec{p}\equiv \vec{p_1}-\vec{p_3}$ and $f_i$ the equilibrium  
Fermi-Dirac distributions for the nucleons. 
 
For the nondegenerate case, in order to solve these phase space 
integrals one follows  
the steps described, for example, in~\cite{raffelt}. The key point is 
the change of the Fermi-Dirac distributions by Maxwell-Boltzmann  
distributions. Neglecting the pion mass we obtain the following expression 
\begin{equation} 
Q_{\nu \bar{\nu}}^{ND} = \frac{65536}{1001\pi^{3/2}} \  
  \frac{\alpha^2_\pi}{m^2_N} \ \frac{1}{\Lambda^4} \ n_n n_p 
  \ \left( \frac{T}{m_N} \right)^{3/2} T^5 .
\end{equation} 
One should consider instead of $m_N$ the effective nucleon mass 
in a dense medium $m_N^*$\cite{raffelt}. The error made neglecting this 
fact is small and opposed to the error made using a vanishing pion mass, so 
that the two errors nearly compensate each other. 
 
In the extreme degenerate case we follow the technique described in  
\cite{friman}; only the contributions of nucleon momenta 
near the Fermi momentum $p_F$ are considered in the phase space integrals
in Eq.~(\ref{phspace}). We find                         
\begin{equation} 
Q_{\nu \bar{\nu}}^D = \frac{31\pi^6}{2970} \ \alpha_\pi^2 \ 
  \frac{1}{\Lambda^4} \ F(p_F/2m_\pi) \ \frac{T^{10}}{m_\pi}, 
\end{equation} 
with  
\begin{equation} 
F(u) \equiv \frac{2}{\pi} \left( \arctan u - \frac{11}{5} \frac{u}{u^2+1}  
  +\frac{26}{15} \frac{u}{(u^2+1)^2} - \frac{8}{15} \frac{u}{(u^2+1)^3}  
  \right). 
\end{equation} 
For $p_F \approx 480$ MeV and $m_\pi = 140$ MeV, $F(u) \approx 0.71$. 
 
The observational constraint $Q_{\nu \bar{\nu}} V_c < 10^{52}$ erg/s gives 
the following bounds 
\begin{eqnarray} 
\Lambda & > & 90 \; \mbox{TeV} \; \; \; \; \; \mbox{ND nucleons}, \\  
\Lambda & > & 150 \; \mbox{TeV} \; \; \; \; \mbox{D nucleons}. 
\end{eqnarray} 
As mentioned before the actual bound on $\Lambda$ falls between these two 
values. 

We have estimated that for scales of the order $\Lambda > 1$ TeV 
the $\nu_R$ are not trapped in the core 
by rescattering or pair absorption due to the contact interactions in 
Eq.~(\ref{nuq}). Since they do not participate in ordinary weak interactions,
once they are produced they leave 
the star and do not contribute to the energy transport inside the core.

To sum up, we have considered contact interactions between quarks
and right-handed neutrinos. We expect the scale $\Lambda$ of such
interactions to be of the same order of magnitude as the
electron-quark interactions, reconsidered recently at the light of HERA 
data. We have shown that they may lead to potential effects
in a supernova. We have restricted the scale of the contact interaction
to be $\Lambda>170$ TeV for nondegenerated nucleons and $\Lambda>250$ 
TeV for degenerated nucleons, when quarks have both axial and vector 
couplings. In the case that quarks have only vector couplings, 
we have evaluated
the production mechanism consisting in neutrino emission from virtual
pions and found the bound $\Lambda>90$ TeV for nondegenerate nucleons
and $\Lambda>150$ TeV for degenerate nucleons. We conclude that in 
models with right-handed (Dirac) neutrinos the scale of contact interactions
should at least be $\Lambda>90$ TeV. 

\vspace{1cm}
{\small We thank the Theoretical Astroparticle Network for support 
under the EEC Contract No. CHRX-CT93-0120 (Direction
Generale 12 COMA). This work has been partially supported by
the CICYT Research Project Nos. AEN95-0815 and AEN95-0882.
R.T. acknowledges the financial support of the Ministerio de
Educaci\'{o}n y Ciencia (Spain).}


\newpage    
\begin{thebibliography}{99}
\bibitem{general}
G. Altarelli et al., hep-ph/9703276;\\
K.S. Babu et al., Phys. Lett. B402, 367 (1997);\\
V. Barger et al., Phys. Lett. B404, 147 (1997);\\
M.C. Gonzalez-Garc\'{\i}a and S.F. Novaes, Phys. Lett. B407, 255 (1997);\\
N. Di Bartolomeo and M. Fabbrichesi, Phys. Lett. B406, 237 (1997);\\
A.N. Nelson, Phys. Rev. Lett. 78, 4159 (1997);\\
W. Buchmuller and D. Wyler, Phys. Lett. B407, 147 (1997);\\
N.G. Deshpande, B. Dutta and X-G He, hep-ph/9705236;\\
K.S. Babu, C. Kolda and J. March-Russell, hep-ph/9705414;\\
F. Caravaglios, hep-ph/9706288;\\
L. Giusti and A. Strumia, hep-ph/9706298;\\
D. Zeppenfeld, hep-ph/9706357;\\
F. Cornet and J. Rico, hep-ph/9707299;\\
G-C Cho, K. Hagiwara, and S. Matsumoto, hep-ph/9707334;\\
V. Barger et al., hep-ph/9707412.
\bibitem{h1}
The H1 coll., C. Adloff et al., Z. Phys. C74, 191 (1997).
\bibitem{zeus}
The ZEUS coll., J. Breitweg et al., Z. Phys. C74, 207 (1997).
\bibitem{PDG}
Particle Data Group, Phys. Rev. D54, 1 (1996).
\bibitem{cornet}
F. Cornet and J. Rico in Ref.~[1].
\bibitem{deshpande}
N.G. Deshpande, B. Dutta, and X-G. He in Ref.~[1].
\bibitem{bw}
W. Buchmuller and D. Wyler in Ref.~[1].
\bibitem{noecker}
M.C. Noecker et al., Phys. Rev. Lett. 61, 310 (1988). 
\bibitem{raffelt} 
G.G. Raffelt, Stars as Laboratories for Fundamental 
Physics, The University of Chicago Press (1996).
\bibitem{friman}
B.L. Friman and O.V. Maxwell, Ap. J. 232, 541 (1979).
\bibitem{brinkmann}
R.P. Brinkman and M.S. Turner, Phys. Rev. D38, 2338 (1988).
\end{thebibliography}
\end{document}